\RequirePackage{rotating}
\documentclass[manuscript]{acmart}

\usepackage{array} 

\usepackage{braket}
\usepackage{multirow}
\usepackage{lscape}

\newcommand{\hide}[1]{}

\AtBeginDocument{%
  \providecommand\BibTeX{{%
    \normalfont B\kern-0.5em{\scshape i\kern-0.25em b}\kern-0.8em\TeX}}}

\newcommand{\bigO}{\ensuremath{\mathcal{O}}}

\usepackage{titlesec}
\titlespacing{\section}{1pt}{2pt plus 2pt minus 2pt}{2pt plus 2pt minus 2pt}
\titlespacing{\subsection}{1pt}{2pt plus 2pt minus 2pt}{2pt plus 2pt minus 2pt}
\titlespacing{\subsubsection}{1pt}{2pt plus 2pt minus 2pt}{2pt plus 2pt minus 2pt}

\setcopyright{none}

\begin{document}

\title{Teaching Algorithm Design: A Literature Review}

\author{Jonathan Liu}
\email{jonliu@uchicago.edu}
\affiliation{
  \institution{University of Chicago}
  \city{Chicago}
  \state{Illinois}
  \country{USA}
}

\author{Seth Poulsen}
\email{seth.poulsen@usu.edu}
\affiliation{
  \institution{Utah State University}
  \city{Logan}
  \state{Utah}
  \country{USA}
}

\author{Erica Goodwin}
\email{egoodwin@uchicago.edu}
\affiliation{
  \institution{University of Chicago}
  \city{Chicago}
  \state{Illinois}
  \country{USA}
}

\author{Hongxuan Chen}
\email{hc10@illinois.edu}
\affiliation{
  \institution{University of Illinois at Urbana-Champaign}
  \city{Urbana}
  \state{Illinois}
  \country{USA}
}

\author{Grace Williams}
\email{gcwill@uchicago.edu}
\affiliation{
  \institution{University of Chicago}
  \city{Chicago}
  \state{Illinois}
  \country{USA}
}

\author{Yael Gertner}
\email{ygertner@illinois.edu}
\affiliation{
  \institution{University of Illinois at Urbana-Champaign}
  \city{Urbana}
  \state{Illinois}
  \country{USA}
}

\author{Diana Franklin}
\email{dmfranklin@uchicago.edu}
\affiliation{
  \institution{University of Chicago}
  \city{Chicago}
  \state{Illinois}
  \country{USA}
}

\begin{abstract}
Algorithm design is a vital skill developed in most undergraduate Computer Science (CS) programs, but few research studies focus on pedagogy related to algorithms coursework. To understand the work that has been done in the area, we present a systematic survey and literature review of CS Education studies. We search for research that is both related to algorithm design (as described by the ACM Curricular Guidelines) and evaluated on undergraduate-level students. Across all papers in the ACM Digital Library prior to August 2023, we only find 94 such papers. 

We first classify these papers by topic, evaluation metric, evaluation methods, and intervention target. Through our classification, we find a broad sparsity of papers which indicates that many open questions remain about teaching algorithm design, with each algorithm topic only being discussed in between 0 and 10 papers. We also note the need for papers using rigorous research methods, as only 38 out of 88 papers presenting quantitative data use statistical tests, and only 15 out of 45 papers presenting qualitative data use a coding scheme. Only 17 papers report controlled trials. In addition, almost all authors only contribute to one publication, an indication that few groups have been able to gain deep insights built on their prior work. 

We then synthesize the results of the existing literature to give insights into what the corpus reveals about how we should teach algorithms. Broadly, we find that much of the literature explores implementing well-established practices, such as active learning or automated assessment, in the algorithms classroom. However, there are algorithms-specific results as well: a number of papers find that students may under-utilize certain algorithmic design techniques, and studies describe a variety of ways to select algorithms problems that increase student engagement and learning. 

The results we present, along with the publicly available set of papers collected, provide a detailed representation of the current corpus of CS Education work related to algorithm design and can orient further research in the area. 
\end{abstract}




\maketitle

\section{Introduction}
The study of algorithms is widely regarded as a critical part of an undergraduate Computer Science (CS) degree \cite{curricular} and has clear practical applications, especially in software-related careers. Furthermore, algorithm skills often play a large part in job interviews for new graduates attempting to obtain a job after completing their computing degree~\cite{behroozi2019hiring}.
Though the design of rudimentary algorithms has been extensively researched in an introductory programming context, most students will take a more in-depth algorithms-focused course later in their degree, as suggested by the ACM curricular guidelines~\cite{curricular}.
Research on algorithm design instruction, as taught in upper-level algorithms courses, is critically underrepresented in the literature. As such, many open questions remain about student experiences and difficulties in learning this material, as well as pedagogical practices for teaching algorithm design in an effective and equitable manner. 

In this paper, we present a systematic survey and review of CS Education literature related to algorithm design education, aiming to answer the following research questions: 

\begin{itemize}
\item \textbf{RQ1:} What algorithm topics are currently represented in the literature, and what kinds of studies are being conducted on these topics?
\item \textbf{RQ2:} What kinds of interventions have been developed to teach algorithm design?
\item \textbf{RQ3:} What knowledge currently exists in the literature about undergraduate algorithm design courses? What are the major takeaways?
\end{itemize}

The main contributions of this work are to:
\begin{itemize}
\item Characterize the current literature with respect to methods, topic, and authorship,
\item Synthesize the existing knowledge about teaching algorithms, and
\item Identify opportunities for future research.
\end{itemize}

\section{Theory and Related Work}
\subsection{Systematic Literature Reviews}
Systematic literature reviews (SLR) are an important part of pushing a research field forward. According to Cooper \cite{Cooper1988Taxonomy}, as a field expands, a synthesis of existing knowledge pertaining to the topic area becomes increasingly necessary to help researchers stay up to date with current developments, especially in topics related to one's primary specialty. Furthermore, according to Kitchenham and Charters \cite{LitReview}, an SLR can help researchers know what is and is not known in the field as they plan for future studies, and can provide a framework through which any new discourse on the topic can contribute. 

Cooper \cite{Cooper1988Taxonomy} also provides a taxonomy of six major characteristics through which SLRs can be classified. We list the six characteristics below, along with the appropriate categories for this review, to help situate our contribution. 
\begin{itemize}
    \item \textbf{Focus:} Our review primarily concerns itself with \textit{research methods} and \textit{research outcomes}.
    \item \textbf{Goal:} Our review aims to \textit{synthesize prior literature} and \textit{identify issues central to the field}. 
    \item \textbf{Perspective:} Our review takes the role of \textit{neutral representation}, aiming to describe what is discussed by the literature without arguing for a specific point of view. 
    \item \textbf{Coverage:} Our review's coverage of papers is \textit{exhaustive with selective citation}. Though we believe our review includes most of the literature on the topic, some of the papers are not cited in this review. That said, a list of all papers reviewed is publicly available (omitted for anonymization purposes). 
    \item \textbf{Organization:} Our review organizes papers \textit{conceptually}, both by topic studied and by results. 
    \item \textbf{Audience:} Our review is intended for \textit{specialized and general researchers}, as well as \textit{practitioners}. CS Education researchers may use this review to understand the topic and situate new research, whereas practitioners who teach algorithms or related courses may use this review to discover interventions or insights for use in their own courses. 
\end{itemize}

\subsection{Literature Reviews in CS Education}
SLR papers have appeared regularly in computing education venues in recent years, with reviews being conducted on topics like CS1~\cite{luxton2018introductory, becker50cs1}, Parsons problems~\cite{ericson2022parsons,du2020review}, undergraduate teaching assistants~\cite{Mirza19UTA}, meta-cognition in programming~\cite{prather2020we}, use of theory in computing education research~\cite{malmi2019in},
and demographic data reporting~\cite{oleson2022decade}. To our knowledge, there have been no literature reviews seeking to understand the state of algorithms education.

Methodologically, two reviews very similar to ours are by Sheard et al. \cite{Sheard09Analysis}, who use Simon's classification scheme to present a detailed quantitative landscape of programming education papers with some important outcomes listed, and by \v{S}v\'{a}bensk\'{y} et al. \cite{svabensky20cyber}, who quantitatively extract and present topics, methods, evaluation, impact, and contributors from cybersecurity education papers, analyzing trends of these characteristics without much focus on findings. These SLRs place a larger emphasis on the focus and approach of existing research in the area. We take a similar approach because the relative sparsity of work means that findings are difficult to synthesize but the methods and topics explored by these papers can still be used to guide future work. 

Topic-wise, the closest systematic reviews are on algorithm visualizations and their effectiveness~\cite{shaffer2007algorithm,shaffer2010algorithm,hundhausen2002meta}, one important subarea of algorithms instruction. While there is some overlap, these reviews cover a largely different set of literature than the literature reviewed in this paper. Both studies cover algorithm visualization of traditional algorithms (e.g. greedy programming). However, algorithm visualization reviews include topics that we do not include, such as using algorithm visualizations to learn $\bigO(n^2)$ sorting algorithms or basic data structure traversals, as they are typically taught prior to algorithms courses \cite{curricular, algsCourses}. In addition, our review covers non-visualization interventions and studies that seek to better understand student experiences in learning algorithms without providing an intervention. In fact, such non-visual areas make up a large proportion of the studies incorporated in this review. 

\section{Methods}
To investigate our research questions, we conducted a systematic literature review, following guidelines from Kitchenham and Charters ~\cite{LitReview} and Randolph \cite{randolph09review}. 

\subsection{Search Strategy}
\label{sec:search-strategy}
We conducted our search on the ACM Full-Text Collection in the ACM Digital Library (DL). To capture our research questions with an appropriate query, we first identified the set of topics that could be considered appropriate for an algorithms course. Within the "Algorithms and Complexity (AL)" knowledge area proposed by the most recent ACM Curricular Guidelines \cite{curricular}, the sections \textit{AL/Algorithmic Strategies} and \textit{AL/Fundamental Data Structures and Algorithms} relate directly to our research questions. To distinguish between data structures topics and algorithms topics, we removed topics that most schools teach prior to their algorithms course as found by Luu et al. \cite{algsCourses} (e.g. Sequential and binary search algorithms). 

For each topic remaining, we designed and validated a query to capture the corpus of literature on the topic. For some topics, this was straightforward:  \textit{Greedy algorithms} was captured by the query ``greedy''. For others, this required more care: for  \textit{Reduction: transform-and-conquer}, ``reduction'' was too common to be a reasonable query, whereas ``transform-and-conquer'' was too specific, but the query \{``algorithm reduction'' OR ``reduction algorithm''\} captures papers relevant to our interests. Included topics and their corresponding queries are shown in Table \ref{tab:topics}.

\begin{table*}[!b]
  \caption{Topics and Queries Used}
  \vspace{-0.1in}
  \label{tab:topics}
  \begin{tabular}{m{1.3in}|m{4.4in}}
    \toprule
    Topic & ACM Digital Library Query\\ 
    \hline
    Branch-and-bound & "branch and bound"\\ [.1em]
    Brute-force algorithms & "brute force" AND algorithm \\ [.1em]
    Depth- and breadth-first traversals & "breadth first search" OR "depth first search" OR "breadth first traversal" OR "depth first traversal" \\ [.1em]
    Divide-and-conquer & "divide and conquer"\\ [.1em]
    Dynamic programming & "dynamic programming"\\ [.1em]
    Greedy algorithms & "greedy"\\ [.1em]
    Heuristics & "heuristic algorithms"\\ [.1em]
    Minimum spanning tree & "minimum spanning tree" OR "prim's algorithm" OR "kruskal's algorithm" \\ [.1em]
    Pattern matching and string/text algorithms & "string algorithm" OR "text algorithm" OR "substring matching" OR "regular expression matching" OR "longest common subsequence"\\ [.1em]
    Recursive backtracking & "recursive backtracking"\\ [.1em]
    Reduction: transform-and-conquer & "algorithm reduction" OR "reduction algorithm"\\ [.1em]
    Representations of graphs & "adjacency list" OR "adjacency matrix" \\ [.1em]
    Shortest-path algorithms & "shortest path algorithm" OR "dijkstra's algorithm" OR "floyd's algorithm" \\ [.1em]
    $O(n\log n)$ Sorting Algorithms & "quick sort" OR "quicksort" OR "heap sort" OR "heapsort" OR "merge sort" OR "mergesort" \\ 
    \bottomrule
  \end{tabular}
\end{table*}

With this set of queries, we conducted three searches within the ACM Digital Library (DL). First, we searched for all \textit{research articles} from venues sponsored by or in collaboration with SIGCSE (366 unique entries). Papers from SIGCSE TS and ITiCSE between 1970 and 2008 are classified as \textit{articles} in the DL instead because they were published in the SIGCSE Bulletin, so we searched these as well (454 entries). Finally, we searched for research articles in TOCE (56 entries). From this corpus, we removed any retracted papers, ACM Bulletin Member Spotlights, or papers without PDFs in the DL. 

We note that this literature search omits venues unaffiliated with the ACM DL. This decision was made in large part due to consistency in search mechanism and capacity constraints. We recognize that venues like the Taylor \& Francis \textit{Journal of Computer Science Education}, the Sage \textit{Journal of Educational Computing Research}, and the CCSC regional conferences, among others, provide important contributions to the CS Education discourse as well, and encourage exploration of these venues for work related to our topic. That said, the set of venues included is consistent with other CS Education literature reviews \cite{Heckman21review}.

\subsection{Study Selection}
We then applied three further exclusion criteria on the remaining 869 papers. First, we aimed to omit papers not relevant to our search, but variation in algorithms course content made this difficult.  For example, only around 60\% of institutions teach Depth-First Search/Breadth-First Search (DFS/BFS) in their algorithms courses \cite{algsCourses}. To ensure that no relevant papers were excluded, we kept any paper that states involvement in an ``algorithms course'' as well as any paper explicitly teaching any of our topics (see Table \ref{tab:topics}). For example, a paper about a CS2 (Data Structures) course in general would be omitted, but a paper that describes teaching DFS/BFS in CS2 would be kept.

Papers were also only kept if they presented some form of data from college-level students. Either quantitative or qualitative data was accepted, but papers that describe an intervention or tool but do not evaluate it were omitted. 

The exclusion criteria were applied by four of the listed authors, who independently filtered a subset of 20 papers, discussed to reach a consensus, and continued. After the authors filtered the second subset, Fleiss' Kappa statistic \cite{fleiss}  was used to compute the inter-rater agreement for four raters, and the authors achieved Fleiss' Kappa $\kappa > 0.87$, so the rest of the papers were processed independently. In the end, 94 papers were identified as relevant to our literature review.

\subsection{Paper Tagging}\label{subsec:scheme}
The authors then developed a set of tags for describing and classifying the relevant papers. 

\subsubsection{Codebook Development}

A Content Analysis \cite{Drisko16content} protocol was used to develop a codebook with appropriate categories for our review. We began with a draft codebook with deductive codes and descriptions that all coders agreed upon. We then iteratively improved the codebook by repeating the following steps:
\begin{enumerate}
    \item Each coder independently tags 10 papers according to the codebook.
    \item Any disagreements in coding are resolved through discussion.
    \item The codebook is adjusted to reflect the new shared understanding of the coders. This may involve adding or removing codes, or adjusting the descriptions.
\end{enumerate}
At the end of the third cycle, no changes were made to the codebook. Fleiss' Kappa statistic was used to calculate reliability with a fixed number of raters on categorical items \cite{fleiss}, and the authors achieved Fleiss' Kappa $\kappa > 0.80$ inter-rater agreement on this batch, so the remaining papers were tagged independently. 

\subsubsection{Final Codebook}

The finalized codebook's codes were split into four major sections, described here. Within each section, the list of possible tags does not exhaustively consider every possibility for a paper but does capture every paper found in our review. All tags can be seen in Table \ref{tab:by_topic}.

\textbf{Topic:} Each paper was tagged by the topic or topics from the ACM curricular guidelines it focused on, and an extra \textit{Not topic-specific} tag was added for papers that studied algorithms pedagogy as a whole without focusing on a specific topic. Our codebook specifies that topic tags are reserved for when the \textit{Results} section of the paper explicitly describes the topic, so that whole-course interventions are tagged as \textit{Not topic-specific} rather than with every individual topic taught in the course. 

\textbf{Evaluation:} This set of tags aims to describe the data presented by the paper. These tags include \textit{Quantitative Data} and \textit{Qualitative Data} as broad categories, but also data gathering methods (e.g. \textit{Survey}, \textit{Interview}) and data analysis methods (e.g. \textit{Statistical Tests}, \textit{Thematic Coding}). Note that papers could and often did contain multiple of these tags, and that some tags even imply others. For example, if a paper was tagged with \textit{Thematic Coding}, then it was also tagged with \textit{Qualitative Data}. 

\textbf{Metric:} This set of tags aims to determine the student metric measured by the paper. We found all papers presented data on at least one of \textit{Performance}, \textit{Affect}, or \textit{Persistence} (whether students remain enrolled in the course). 

\textbf{Intervention:} This set of tags aims to capture information about the intervention developed in the paper, if applicable. We categorized interventions based on the aspect of the course targeted, and introduced a \textit{Tools} tag because some interventions developed a tool without directly affecting the course. These were not mutually exclusive --- for example, a tool may be developed to help students during discussion. A full list and description of intervention tags can be found in Table \ref{tab:tags}.

\begin{table*}[h] 
  \caption{Tagging Scheme to Characterize Interventions}
  \vspace{-0.1in}
  \label{tab:tags}
  \begin{tabular}{m{1.1in}|m{4.5in}}
    \toprule
    Tag & Description \\
    \midrule
    None & A study that does not change the way students learn the material, and instead investigates the status quo, e.g. misconceptions or student behavior. \\ [.4em]
    Tools & A study that develops and/or evaluates a tool that a student can interact and interface with, especially a software tool, that aids with student learning. \\ [.4em]
    \midrule
    Course Policy & A study that changes course logistics unrelated to content (e.g. late policy or collaboration policy). \\ [.4em]
    Content Presentation & A study that changes how the instructor presents course material. \\ [.4em]
    Discussion/Lab & A study that changes staff-facilitated problem-solving sessions (e.g. labs and group work).\\ [.4em]
    Student Work & A study that changes non-staff-facilitated work  (e.g. homework or projects). \\ [.4em]
    Exams & A study that changes how students are evaluated in test settings. \\ [.4em]
    \bottomrule
  \end{tabular}
  \vspace{-0.1in}
\end{table*}

\subsection{Results Consolidation}
To identify major takeaways from the corpus, the papers were split based on whether they presented an intervention. For papers containing interventions, the first and third authors read each paper, recording the results of the paper as well as any features of the intervention explicitly described in the paper as beneficial. These results and features were consolidated and inductively coded by the authors. Papers not containing an intervention were instead inductively coded based on focus. The final codes can be found as the headers in Section \ref{sec:results}.

\section{Corpus Survey}
The results of the tagging can be found in Table \ref{tab:by_topic}. Note that a single paper can be tagged in multiple categories and topics, so it can be represented in multiple rows and columns. The last row contains a summary of the results. The fully tagged corpus can be found at \url{https://doi.org/10.7910/DVN/KTR2ZH}.

We begin by presenting a broad survey of the methodology used in the corpus. We then drill down to answer more specific questions about topic coverage, changes over time, ``rigor'' (which we define later), and the venues and authors that publish the most. Findings are accompanied by a discussion about implications and future steps.

\begin{table*}[h]
  \caption{Tags by Topic. Note that some papers cover multiple topics, and contribute to each corresponding row in the table.}
  \vspace{-0.1in}
  \label{tab:by_topic}
  \begin{tabular}{m{1.in}c cccccccc ccc ccccccc}
    \toprule
    & & \multicolumn{8}{c}{Evaluation} & \multicolumn{3}{c}{Metric} & \multicolumn{7}{c}{Intervention}\\
    \cmidrule(r){3-10} \cmidrule(lr){11-13} \cmidrule(l){14-20} 
    Topic & Total & \rotatebox[origin=ct]{90}{Quantitative Data} & \rotatebox[origin=ct]{90}{Survey} & \rotatebox[origin=ct]{90}{Scores/Grades} & \rotatebox[origin=ct]{90}{Statistical Tests} & \rotatebox[origin=ct]{90}{Controlled Trial} & \rotatebox[origin=ct]{90}{Qualitative Data} & \rotatebox[origin=ct]{90}{Interview} & \rotatebox[origin=ct]{90}{Thematic Coding} & \rotatebox[origin=ct]{90}{Performance} & \rotatebox[origin=ct]{90}{Affect} & \rotatebox[origin=ct]{90}{Persistence} & \rotatebox[origin=ct]{90}{None} & \rotatebox[origin=ct]{90}{Tools} & \rotatebox[origin=ct]{90}{Course Policy} & \rotatebox[origin=ct]{90}{ Content Presentation} & \rotatebox[origin=ct]{90}{Discussion/Lab} & \rotatebox[origin=ct]{90}{Student Work} & \rotatebox[origin=ct]{90}{Exams} \\ 
    \hline
    Branch-and-bound & 1 & 1 & 0 & 1 & 0 & 0 & 1 & 0 & 1 & 1 & 0 & 0 & 1 & 0 & 0 & 0 & 0 & 0 & 0\\ [.3em]
    Brute-force \\Algorithms & 2 & 2 & 1 & 2 & 2 & 2 & 0 & 0 & 0 & 2 & 1 & 1 & 0 & 2 & 0 & 0 & 1 & 1 & 0 \\[.3em]
    Depth- and Breadth-First Traversals & 7 & 6	& 5	& 4	& 3 & 3	& 3	& 1	& 1	& 4	& 5	& 0	& 0	& 3	& 0	& 2	& 1 & 3 & 0 \\ [.3em]
    Divide-and-Conquer & 3 & 3 & 1 & 3 & 2 & 1 & 2 & 1 & 0	& 3	& 0	& 0	& 2	& 1	& 0	& 0	& 1	& 0	& 0 \\ [.3em]
    Dynamic \\Programming & 10 & 10 & 6 & 8 & 5 & 3 & 5 & 2 & 1 & 8 & 5 & 1 & 3 & 4 & 0 & 1 & 1 & 3 & 0 \\ [.3em]
    Greedy Algorithms & 8 & 8 & 4 & 6 & 0 & 0 & 6 & 2 & 5 & 6 & 4 & 0 & 2 & 3 & 0 & 3 & 2 & 2 & 0 \\ [.3em]
    Heuristics & 3 & 3 & 0 & 3 & 0 & 0 & 2 & 0 & 2 & 3 & 0 & 0 & 1 & 1 & 0 & 0 & 0 & 2 & 0 \\ [.3em]
    Minimum Spanning Trees & 5 & 5 & 4 & 3 & 1 & 1 & 2 & 0 & 0 & 3 & 3 & 1 & 0 & 3 & 0 & 0 & 1 & 4 & 1 \\ [.3em]
    Pattern Matching and String/Text Algorithms & 1 & 1	& 1	& 0	& 0	& 0	& 1	& 0	& 0	& 0	& 1	& 0	& 0	& 0	& 0	& 0	& 0	& 1	& 0 \\ [.3em]
    Recursive \\Backtracking & 3 & 3 & 2	& 3	& 1	& 1	& 2	& 0	& 1	& 3	& 2	& 0	& 1 & 2 & 0 & 1 & 1 & 0 & 0 \\ [.3em]
    Reductions & 6 & 5 & 3 & 6 & 1 & 0 & 4 & 1 & 2 & 6 & 2 & 0 & 3 & 0 & 1 & 2 & 0 & 2 & 0 \\ [.3em]
    Representations of Graphs & 0 & 0 & 0 & 0 & 0 & 0 & 0 & 0 & 0	& 0	& 0 & 0 & 0 &	0 & 0 &	0 &	0 &	0 & 0\\ [.3em]
    Shortest-path \\Algorithms & 6 & 6 & 4 & 5 & 2 & 2 & 1 & 0 & 0 & 5 & 4 & 1 & 0 & 6 & 0 & 1 & 1 & 1 & 0 \\ [.3em]
    $\bigO(n\log n)$ Sorting \\Algorithms & 10 & 9 & 7 & 6 & 5 & 2 & 5 & 1 & 0 & 6 & 9 & 1 & 1 & 4 & 0 & 2 & 2 & 4 & 0 \\ 
    \midrule 
    Total (any topic) & 46 & 43 & 28 & 35 & 15 & 10 & 23 & 6 & 8 & 35 & 27 & 2 & 9 & 20 & 1 & 10 & 9 & 15 & 1 \\
    Not topic-specific & 48	& 45 & 27 & 34 & 23 & 7	& 22 & 6 & 7 & 34 & 30 & 2 & 11 & 16 & 11 & 9 & 7 & 16 & 2 \\
    \midrule
    Total & 94 & 88 & 54 & 69 & 38 & 17 & 45 & 12 & 15 & 69 & 57 & 4 & 20 & 36 & 12 & 19 & 16 & 31 & 3 \\
    \bottomrule
  \end{tabular}
  \vspace{-0.1in}
\end{table*}

\subsection{Corpus Methodology}
Nearly every paper (88/94) meeting our selection criteria presents quantitative data, and 73\% (69/94) of papers use student scores and grades. Notably, around 40\% (38/94) of the papers run statistical tests, and only around 18\% (17/94) run controlled trials. On the other hand, only half of the papers use qualitative data. Thematic coding, found in only 16\% (15/94) of papers, was done for interview responses (e.g. \cite{Toma18SelfEfficacy}), student work (e.g. \cite{velazquez12refinement}), and survey responses (e.g. \cite{weber23altgrading}). Both quantitative and qualitative data are presented in 41\% (39/94) of the papers.

In terms of evaluation metrics, 73\% (69/94) of papers use student performance, 61\% (57/94) use affective measures, and 34\% (32/94) of papers use both. Of the four papers measuring persistence \cite{Garcia-Mateos09Judging, deb2017creating, Lin21PDL, Coffey13Integ}, all four measure performance as well, and three measure affect, so it may be worthwhile to conduct studies more centered around persistence in the algorithms class. 

Most papers (79\%, 74/94) present interventions, and around 40\% (36/94) center around a tool. For example, five papers present a strategy for implementing active learning in the classroom via interactive  question-and-answer tools \cite{pargas06reducing, anderson07classroompresenter, kurtz14active, Phan18Brownies, deb18MRS}. On the other hand, around a fifth (20/94) of the papers surveyed do not develop an intervention. Many of these papers analyze student problem-solving strategies when presented with algorithm design problems, and a number of others study topic-specific misconceptions.

We only find four interventions related to student examinations. Given the broad applicability of algorithms to test questions and the effect that question design and test logistics can have on exam validity, we expect research on this subject to be potentially useful.

\subsection{Representation of Topics}
\label{sec:topics}
Of the 94 papers included in our literature review, nearly half focus on a specific topic or set of topics, whereas the other half present results that do not pertain to any particular topic, such as studies about broad problem-solving skills or policy-based interventions.

Most topic-specific papers only focus on one topic, though ten papers focus on two topics, three were tagged with three topics each \cite{deb2017creating, Taylor09Predic, Brown22Ethics}, and one paper was tagged with four (\textit{Branch-and-Bound}, \textit{Greedy Algorithms}, \textit{Heuristics}, \textit{Recursive Backtracking})  \cite{Vel2019Misconceptions}. However, many areas are still understudied. In this section, we begin by discussing trends in individual topics, and then compare topic-specific papers with those that are not topic-specific.

\subsubsection{Topics with More Coverage}
At 10 papers, $\mathbf{\bigO(n \, log \, n)}$ \textbf{Sorting Algorithms} are the topic most covered in the literature, though we expect that $\bigO(n^2)$ algorithms are studied far more thoroughly. Most of these papers leverage sorting as a familiar task to develop engaging active-learning interventions for introducing faster sorting algorithms, including through games \cite{Hakulinen11Serious, Hosseini19Learning} and exploratory activities \cite{Rasala94Animation, Rebelsky05NewScience, mccann04column}, though a couple instead use assessment results to provide insight into common misconceptions \cite{winters06Closing, taherkhani12sort}. 

Research about the topic of \textbf{Dynamic Programming} has taken a relatively wide variety of approaches. Two papers contribute interesting course projects that motivate the use of Dynamic Programming through various problems, one from a centuries-old mathematical text \cite{pengelley06project} and the other from the modern world \cite{Bird06Search}. Tools have also been developed to help scaffold different steps of solving DP problems, specifically interpreting the problem constraints \cite{Lin21PDL}, visualizing subproblem recurrences \cite{velazquez16multiple, Velazquez16Systematic}, and writing a proof \cite{Xia23CFG}. Research has been conducted \cite{danielsiek12misconceptions, Paul13hunting} and reproduced \cite{DPmisconceptions} investigating common roadblocks in solving DP problems as well, finding that students either don't recognize that DP is the correct approach, fail to split the problem into appropriate subproblems, or have trouble identifying the correct recursion of subproblems. Finally, Enstr\"{o}m and Kann \cite{enstrom17iterating} find that teaching the DP process as separate steps can increase student self-efficacy . 

On the other hand, \textbf{Greedy Algorithms} has seen 8 papers but markedly less breadth, with one paper about misconceptions \cite{Vel2019Misconceptions}, one about collective argumentation \cite{Kallia22Collective}, and the other six about interactive learning interventions. Though five provide thematic coding of student responses, none provide statistical tests in their evaluations, limiting our understanding of intervention effectiveness.

\subsubsection{Topics with Less Coverage}
We found no papers related to \textbf{Graph Representations} and one related to \textbf{Pattern-Matching and String/Text Algorithms} \cite{Bird06Search}, though the latter may be partially captured in research about Dynamic Programming and Greedy Algorithms. Despite \textbf{Divide-and-Conquer}, \textbf{Minimum Spanning Trees (MST)}, and \textbf{Shortest-path Algorithms} being taught in more than 80\% of algorithms courses \cite{algsCourses}, the topics are only studied by 3, 5, and 6 papers, respectively. Of the three Divide-and-Conquer papers, two focus on misconception detection in a wide set of topics that includes Divide-and-Conquer \cite{danielsiek12misconceptions, Paul13hunting}, while the third involves a tool for learning in the classroom \cite{Phan18Brownies}, so questions remain about presenting, testing, and assigning work on the topic. Four of the five MST papers relate to the development of student assignments \cite{deb2017creating, Taylor09Predic, erkan18mst, Stasko97Animations}, and the other paper studies exam questions \cite{Gaber23Assessments}, leaving the presentation of the material or common student misconceptions unexplored. Finally, all six Shortest-path papers are tool-based interventions, leaving plenty of room for exploration of misconceptions or pedagogical strategies.

\subsubsection{Papers Without Specific Topics}
Within non-topic-specific papers (half of our set), there is considerable variation in the extent to which the knowledge extracted applies specifically to algorithms courses. For example, some papers test interventions in an algorithms course but could have reasonably tested them in any CS course (e.g. \cite{belleville20invert, pargas06reducing}), some studies include algorithms courses alongside other CS courses in their data set (e.g. \cite{deb18MRS, bennedssen08abstraction}), and some studies are topical to algorithms in particular (e.g. \cite{danielsiek17instrument, collberg04Algo}). 

For the most part, these non-specific studies mirror the topic-specific papers in terms of evaluation and intervention methods used. The most notable difference is in \textit{Course Policy}, where all but one paper focuses on the course as a whole rather than individual topics. This is unsurprising, as we expect course policy changes to affect all topics in a course equally, though Crescenzi et al. \cite{crescenzi13theory} study a sizeable modification to course structure targeted at the course's NP-completeness unit. 

That said, the studies are still notably different in their aims. For example, while most topic-specific papers without interventions focus on identifying misconceptions for the topic, papers without specific topics instead focus on broader measurements like skill development \cite{ginat01metacog}, learning outcomes \cite{bennedssen08abstraction}, and even student affect \cite{danielsiek17instrument}. Likewise, while most content-specific tools presented are related to algorithm visualization, papers without specific topics feature not only visualization tools (e.g. \cite{naps00JHave, lee11toward}) but also tools for facilitating improvements in grading \cite{Garcia-Mateos09Judging}, lectures \cite{pargas06reducing}, and assignments \cite{marshall06moving}.

\subsection{Topics over Time}
\begin{figure}[ht]
  \centering
  \includegraphics[width=0.7\linewidth]{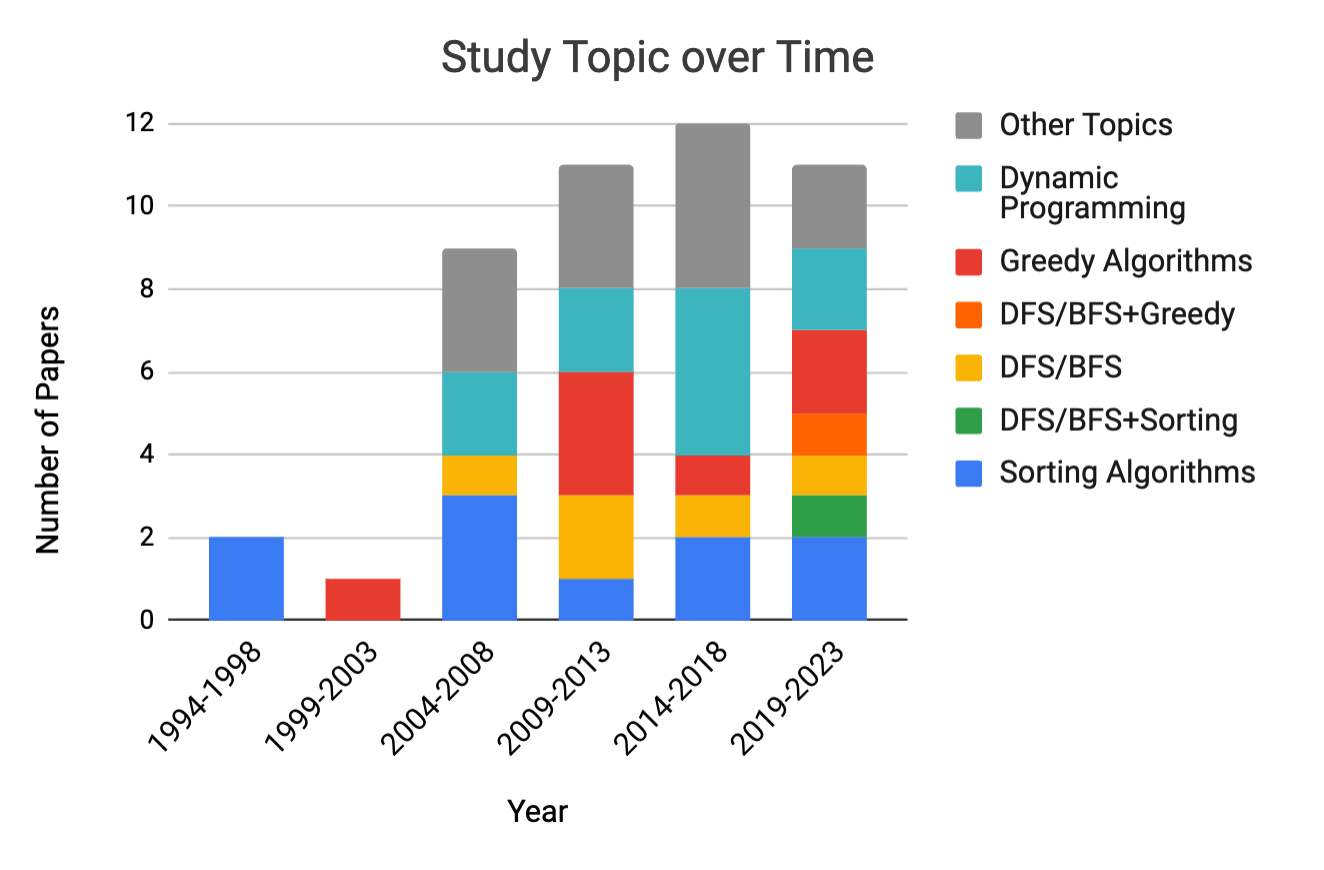}
  \caption{Topic of topic-specific papers, by year}
  \label{fig:topic_by_year}
\end{figure}

In Figure \ref{fig:topic_by_year}, we display the distribution of topics covered by year. Sorting Algorithms dominate the topic-specific literature early on, representing the only topic in 1994-1998 and 5 of 12 between 1994-2008. In every period, at least two-thirds of the papers within our study relate to either Sorting, DFS/BFS, Greedy Algorithms, or Dynamic Programming. 

\subsection{Rigor over Time}
\begin{figure}[ht]
  \centering
  \includegraphics[width=0.7\linewidth]{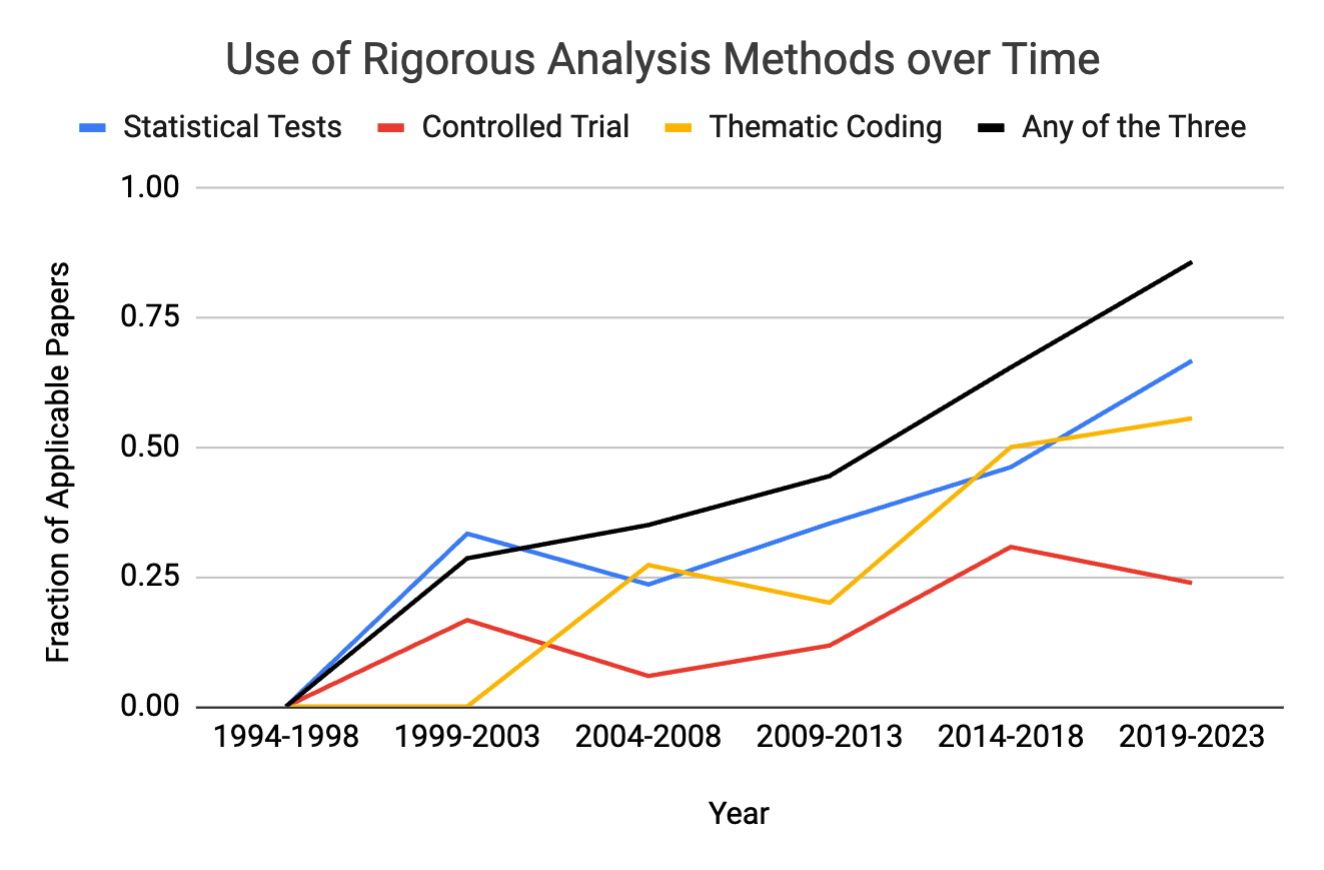}
  \caption{Percentage of papers with statistical tests (quantitative), controlled trials (quantitative), thematic coding (qualitative), and at least one of the above}
  \label{fig:rigor}
\end{figure}

Scientific rigor is by nature tough to quantify, as the best practices for any paper depend on the specific research questions. For papers with quantitative data, the use of \textit{statistical tests} or \textit{controlled trials} serves as our proxy for a metric of rigor \cite{Sanders19Stats}. Similarly, for papers presenting qualitative data, we use \textit{thematic coding}. Note that the papers with no evaluation or student data were already filtered out of our review. Figure \ref{fig:rigor} shows the percentage of each type of study, presented in 5-year bands. 

We see no ``rigorous'' studies initially, but there is an encouraging upward trend in the usage of rigorous procedures in algorithms papers as the field matures. In the most recent period, 2019-2023, 56\% of qualitative papers use thematic coding, and two thirds of quantitative papers use either statistical tests or controlled tests.

\subsection{Where and Who are Studies Coming From?}
\begin{figure}[ht]
  \centering
  \includegraphics[width=0.7\linewidth]{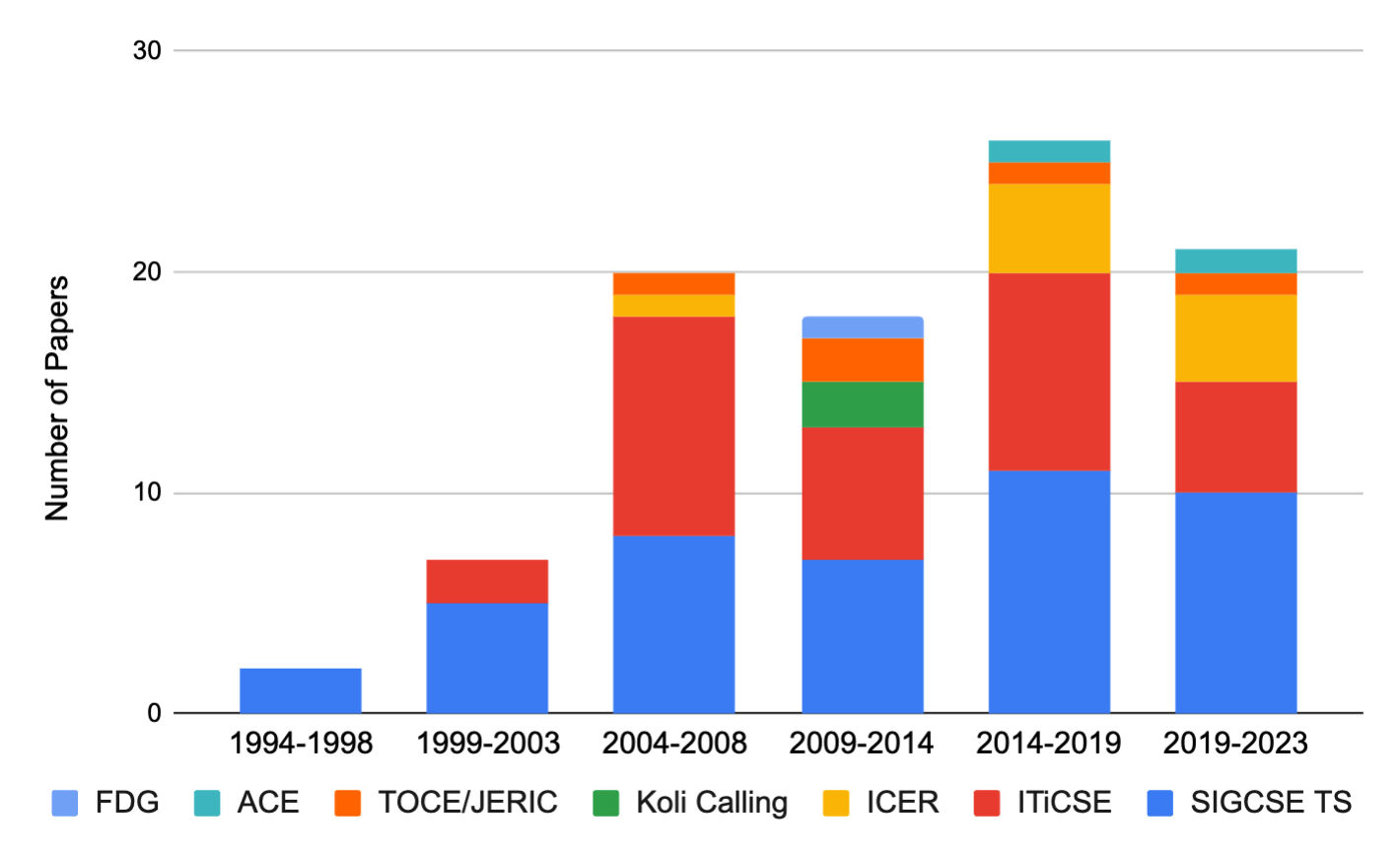}
  \caption{Papers per venue, by year}
  \label{fig:conf}
\end{figure}

Our search consisted of all publications sponsored by or in collaboration with SIGCSE, as well as TOCE. The distributions of papers per conference, split by year, can be found in Figure \ref{fig:conf}.

As expected, SIGCSE's largest conferences, the Technical Symposium and ITiCSE, account for a vast majority of the papers (44 and 33, respectively) in our literature review, with ICER contributing more recently. We found five relevant papers from TOCE (and its previous moniker JERIC) as well.  

We also seek to understand whether much of the work is accomplished by just a few groups, or whether it is widespread in the community. The tally of papers per author can be found in Table \ref{tab:authors}.

\begin{table}[h]
    \vspace{-0.1in}
    \centering
    \caption{Number of Papers per Author. Asterisks indicate frequent collaborators.}
    \vspace{-0.1in}
    \begin{tabular}{cc}
        \toprule
        Author & Number of Publications \\
        \cmidrule(lr){1-1} \cmidrule(lr){2-2}
        J. \'{A}ngel Vel\'{a}zquez-Iturbide* & 7 \\ 
        Jan Vahrenhold** & 5 \\ 
        David Ginat & 4  \\ 
        Antonio P\'{e}rez-Carrasco* & 4 \\
        Michal Armoni & 3 \\
        Holger Danielsiek** & 3 \\\relax
        [12 authors] & 2 \\\relax
        [183 authors] & 1 \\
        \bottomrule
    \end{tabular}
    \label{tab:authors}
    \vspace{-1em}
\end{table}

We can see that the most active groups are two pairs of collaborators. Vel\'{a}zquez-Iturbide / P\'{e}rez-Carrasco have studied optimization algorithms and developed visualization tools to assist with a variety of algorithm paradigms (e.g. \cite{velazquez12refinement, velazquez16multiple, Vel2019Misconceptions, Velazquez16Systematic}) and  Vahrenhold / Danielsiek have explored instruments for measuring student affect and success in algorithms courses (e.g. \cite{danielsiek12misconceptions, danielsiek17instrument, Paul13hunting, Toma18SelfEfficacy}). In addition, David Ginat has written four papers discussing metacognitive strategies for algorithmic problem-solving (e.g. \cite{ginat01metacog, Ginat17Levels}), and Michal Armoni has conducted a series of studies on reductive thinking \cite{armoni09reduction}. The vast majority of authors publish only one or two papers (195 authors), limiting the depth of contributions that might be made. 

\section{Corpus Takeaways}
\label{sec:results}
In this section, we go beyond the numbers to provide insights into the major takeaways from the corpus as a whole. Subsections are broadly sorted in order of decreasing specificity to algorithms courses. 

\subsection{Student Approaches to Algorithm Design Problems} \label{sec:approaches}
Though the skill of algorithm design is somewhat nebulous, there is general consensus that it is closely tied to abstraction ability. Bennedsen and Caspersen \cite{bennedssen08abstraction} examine 10 courses mandatory for the CS degree and find that abstraction ability correlates with performance in only the algorithms course. Think-alouds have been run to identify varied extents to which students utilize abstraction to solve problems \cite{Ginat17Levels, Izu17reason}.

More specific algorithm design skills have been studied as well, with the general finding that students under-utilize them even after having taken an algorithms course. In problems where recursion \cite{Ginat04Recursion} and reduction \cite{armoni09reduction} would have been helpful, students understand the concepts but underestimate their utility in general problem-solving contexts. On the other hand, students demonstrate an overreliance on intuitive greedy approaches, even in the face of counterexamples \cite{Ginat03GreedyTrap}.

At the topic level, a limited number of papers explore common misconceptions in algorithms topics. Danielsiek et al. \cite{danielsiek12misconceptions} use both student work and thinkalouds to cover a wide variety of topics, but other studies have focused on more specific topics like dynamic programming \cite{DPmisconceptions} and optimization algorithms \cite{Vel2019Misconceptions}.

\subsection{Problem Selection}
We find a recurring theme that intentional problem selection can be leveraged to motivate students. The broad applicability of algorithms lends itself nicely to problems tied to the real world, which can give students both familiarity with the problem setting and motivation to solve it. For example, linking shortest path algorithms to GPS navigation is a setting in which multiple papers demonstrate that students enjoy and engage more with the real-world map \cite{holdsworth09GPS, teresco18map}. Mehta et al. \cite{Mehta12Groups} presented bipartite matching in the setting of forming balanced class project groups, resulting in both more successful groupings and an increased understanding of bipartite matching. Problems can also tie into the real world through the researcher: Pengelley et al. \cite{pengelley06project} presented a Dynamic Programming problem posed by a mathematician's letter from 1838, and Wirth and Bertolacci \cite{wirth06research} presented a problem from the instructor's research, both finding that the humanized context motivated students. 

Problems with a variety of solutions can also be particularly helpful. In multiple studies, students indicate that writing both efficient and inefficient solutions to problems and comparing their runtimes on real inputs provided more motivation for designing efficient algorithms than theoretical runtime analyses \cite{mccann04column, Coffey13Integ}. Similarly, studies have designed projects asking students to compare BFS against DFS \cite{erkan08image}, Prim's algorithm against Kruskal's algorithm \cite{erkan18mst}, and several different sorting \cite{Rasala94Animation} and matching \cite{Lucas15Matching} algorithms, finding in all cases that students describe the comparisons as valuable and insightful. In each of these projects, the comparison is facilitated in part by a visualization of either the evaluation or output of the algorithms.

\subsection{Algorithm Visualization}
In our data set, we identified 15 studies about algorithms visualizations. Broadly, students state that they appreciate visualizations, though few studies experimentally measure learning outcomes. Of these papers, three found that introducing visualizations improves student learning \cite{velazquez13experimental, deb2017creating, farghally17visuali}, though only one uses statistical tests. On the other hand, Jarc et al. \cite{jarc2000benefits} find that introducing visualizations provides no statistically significant learning outcomes, and Vel\'{a}zquez-Iturbide and P\'{e}rez-Carrasco \cite{Velazquez16Systematic} find that students are more efficient but do not perform better. Taylor et al. \cite{Taylor09Predic} find that student learning significantly improves when the visualization forces students to take an active, predictive role. For a more thorough review on algorithms visualizations that is not specific to algorithms courses, we recommend Hundhausen et al.'s work \cite{hundhausen2002meta}.

\subsection{Coding Practice}
Though around a quarter of algorithms courses have no programming component \cite{algsCourses}, research has found that algorithmic coding assignments increase student engagement \cite{Garcia-Mateos09Judging} and helps students develop their algorithmic design \cite{izu18unstructured} and runtime analysis \cite{coore19facili} skills. Online coding platforms allow for a public leaderboard of coding problem scores, which some studies have used in an attempt to motivate students \cite{coore19facili, farnqvist16competition} with mixed results: Coore and Fokum \cite{coore19facili} observed that the strongest students worked hard to maintain their ranking, but other students were demotivated when the top scorers had yet to solve a problem.

\subsection{Assessment Policy}
\label{sec:policy}
The use of algorithmic coding problems permits automated assessment, which students find fair and constructive \cite{farnqvist16competition, weber23altgrading}, though additional human interpretation of automated feedback may be helpful \cite{leite20human}. Weber \cite{weber23altgrading} finds promise in standards-based grading for algorithms courses, along with allowing students to re-submit work as they continue to master course concepts. While students did continue to improve on past assignments throughout the course, some took advantage of the chance to procrastinate work until the end of the term.

Two papers investigate the utility of two-stage quizzes (consisting of an individual and a group phase), finding that team quizzes increase students’ perceived and measured learning gains \cite{belleville20invert,ham21teamquiz}. Gaber et al. \cite{Gaber23Assessments} make the case for repeating previously-seen problems on exams, showing that these questions are still considerably challenging for students yet perceived as fair.

\subsection{Active Learning}
Research on algorithms-based active learning exercises in lecture corroborates findings in other settings that students enjoy exploring a problem space with their peers \cite{belleville20invert, vandegrift17pogil, Toma18SelfEfficacy}, feel an increased sense of engagement in class \cite{anderson07classroompresenter, gehringer09studentgenerated}, and report learning both course material and collaboration skills \cite{vandegrift17pogil}. Additionally, incorporating in-class interaction into historically lecture-based courses led to higher instructor satisfaction in some cases \cite{hosseini19areyougame, boucheztichadou18problemsolving}. 

Most studies incorporate active learning by having students work through problems in class, whether exploring a problem instance \cite{anderson07classroompresenter, kurtz14active}, coding an algorithmic solution \cite{Phan18Brownies}, or performing algorithm analysis \cite{pargas06reducing, deibel05team}. Some of these activities were equipped with auto-grading as well, which Deb et al. \cite{deb18MRS} find improved student retention of course material. To dissuade a "guess-and-check" approach while using an auto-grader, Kurtz et al. \cite{kurtz14active} capped grades based on the number of submissions. 

Game-based exercises have also shown promising results, with research finding that these activities can drive participation from students who are typically silent \cite{hosseini19areyougame, Hosseini19Learning}. Some of the algorithms-based games developed include a sorting algorithm design contest \cite{Hosseini19Learning}, a matching game in which students choose algorithms for given criteria, and a draw-and-guess game in which students illustrate course content \cite{Hakulinen11Serious}. 
 
However, despite the success of interactive in-class activities, facilitating productive group work may require further attention: in a study on communication behaviors during a collaborative algorithmic problem solving session, Kallia et al. \cite{Kallia22Collective} observed that first-year undergraduates struggled to build off of each others’ ideas while MSc students were overly dismissive of others' ideas. Two papers find success using a variety of intentional group-formation strategies to encourage productive collaboration, making groups based on learning style, prior student knowledge, or mixed instructor/student input \cite{deibel05team, Mehta12Groups}. 

\subsection{Psychological Factors}
\label{sec:factors}
Little is known about psychological factors in algorithms courses, but a series of papers by Danielsiek, Toma, and Vahrenhold develop and validate an algorithms-specific instrument to measure self-efficacy \cite{danielsiek17instrument} and use it along with other instruments to measure the cognitive load and perceived benefit of different collaborative lab assignments \cite{Toma18SelfEfficacy}. Their work finds that collaborative labs are perceived as effective and inclusive, and demonstrate the potential for using measurements of various emotional responses to guide intervention design.

\section{Limitations}
Though we designed our search query to capture papers related to algorithm design, some papers may have been missed. First, if a paper does not use our search terms, our query may not capture it. Furthermore, our search was limited to venues sponsored by or in collaboration with SIGCSE in the ACM DL, so research published outside of these venues was not included (e.g. the Taylor \& Francis journal \textit{Computer Science Education}), though we expect to have captured most of the work in the area. (See Section \ref{sec:search-strategy}.)

The variability in course offerings across institutions also may have resulted in some omitted papers. Topics that we consider to be relevant to algorithm design are sometimes taught in prerequisite courses like CS2 \cite{algsCourses}, but we may not be able to tell based on the paper. As such, knowledge about teaching these topics in other courses is missed if the paper does not explicitly mention our topic of interest. 

As with any qualitative coding endeavor, we also note that misinterpretations of papers in the review are possible. However, we hope that the content analysis procedure, with many rounds of codebook refinement and discussions about potentially different interpretations of the codebook, mitigated the likelihood of these outcomes.

\section{Suggested Future Work}
As discussed, the sparsity of research on algorithms education suggests many areas for future work. Section \ref{sec:topics} discusses potential areas of exploration at a more fine-grained level.

At a broader level, our review finds that research in the area relies heavily on quantitative data, especially measuring student scores or grades, but relatively few studies use statistical tests and even fewer run controlled trials. Though we acknowledge the difficulty of running controlled trials in an educational setting, more rigorous approaches to this data would be valuable to develop a better understanding of the effectiveness of various interventions.

At the same time, this reliance on quantitative data is accompanied by a dearth of qualitative data, especially about psychological factors like sense of belonging or self-efficacy. Rigorous explorations of student experiences in these classes would be incredibly useful for not only determining the extent to which the mathematical content of algorithms courses affects psychological factors but also orienting future research around mitigating impostor syndrome, stereotype threat, or other discovered barriers. 

While in this review we were able to cover all topics from the ACM Curricular Guidelines sections \textit{AL/Algorithmic Strategies} and \textit{AL/Fundamental Data Structures and Algorithms} which are usually covered in an algorithms course, we were not able to cover topics listed in the related sections \textit{AL/Basic Automata Computability and Complexity} about computing theory and \textit{AL/Basic Analysis} about algorithmic analysis. These topics are also required learning for CS majors according to the ACM Curricular Guidelines, though they are less frequently covered in algorithms courses~\cite{algsCourses}. Future reviews could use these as topics of focus.

\section{Conclusion}
In an effort to synthesize research related to algorithm design education, we categorized all papers in the ACM Digital Library containing search terms for algorithm design topics covered in algorithms courses, guided by the ACM curricular guidelines and surveys of algorithms course instructors~\cite{curricular,algsCourses}. We found that algorithms research is a growing field, with more rigorous methods being applied over time. Studies have generally agreed that active learning interventions and interesting problem-selection can contribute to both learning and motivation, corroborating research in the broader CS Education space. However, even among the most studied areas in algorithm design, there are still large gaps in the literature. For some topics, there is adequate literature on student misconceptions, but concrete instructional interventions are lacking. In others, researchers have created interventions, but have not yet rigorously tested their efficacy using controlled trials. Exams and Psychological Factors stand out as important areas with very little research at all. We hope this review can help guide the future of algorithm design education.



\bibliographystyle{ACM-Reference-Format}
\bibliography{references}

\end{document}